\documentclass[usenatbib,usegraphicx]{mn2e}
\def\lsim{\raise0.3ex\hbox{$<$\kern-0.75em\raise-1.1ex\hbox{$\sim$}}}
\def\gsim{\raise0.3ex\hbox{$>$\kern-0.75em\raise-1.1ex\hbox{$\sim$}}}

\title{Constraints on the Star Formation Rate from
         Supernova Relic Neutrino Observations}
\author[M. Fukugita and M. Kawasaki]
         {M. Fukugita$^{1}$ and M. Kawasaki$^2$\\
          $^1$Institute for Cosmic Ray Research, University of Tokyo,
            Kashiwa 277-8582, Japan\\
         $^2$Research Center for the Early Universe,
         Graduate School of Science, University of Tokyo,
         Tokyo 113-0033, Japan}
\date{Accepted  Received }
\pagerange{\pageref{firstpage}--\pageref{lastpage}}
\pubyear{2002}

\begin{document}

\label{firstpage}

\maketitle

\begin{abstract}
      We discuss the implication of the observation of supernova relic
      neutrinos on the study of the star formation rate (SFR) in 
galaxies.
      The limit recently obtained at
      Super-Kamiokande (SK) is already marginally significant:
      The SFR we derived  $\psi(t_0)<0.040M_{\odot}$yr$^{-1}$Mpc$^{-3}$
      (at a 90\% CL) is about twice the SFR estimated
      from radio observations,
      and five times the rate from H$\alpha$ allowing for uncertainties
      in the model supernova neutrino flux.

\end{abstract}


\section{Introduction}

Neutrinos emitted from stellar core collapse fill the universe as a
diffuse background radiation.  The feasibility for the detection of
these neutrinos has been considered by many authors
(\citealt{Bisnovatyi}; \citealt{Krauss}; \citealt{Totani};
\citealt{Malaney}; \citealt{Hartmann}; \citealt{Kaplinghat};
\citealt{Ando}).  While estimates of the expected neutrino flux depend
much upon authors, the authors are generally negative regarding the
feasibility of their detection. The problem is large backgrounds from
solar and reactor (anti)neutrinos at low energies and atmospheric
neutrinos at high energies; a possible window in between (say, the
neutrino energy $E_\nu=$15-50 MeV) is masked by a large background from
electrons produced by decay of low energy muons that escape detection in
the water $\check{\rm C}$erenkov detector~\citep{Zhang}. The decayed
electron spectrum from muons, however, is precisely known, and the
Super-Kamiokande (SK) group has demonstrated that this background can be
subtracted \citep{Totsuka}. The limit derived on the neutrino event rate
in the 18-50 MeV range is close to the value indicated by some
optimistic estimates of the supernova relic neutrino flux, which
encourages us to scrutinize the problem.

A major uncertainty in the calculation of the supernova relic neutrino
flux is in the star formation rate (SFR) and its evolution towards the
past. The work with high redshift galaxies over the last five years,
however, has provided us with significant insight on the evolution of
the global SFR. The estimates include the use of UV emissivity
(\citealt{Madau}, \citealt{Lilly}, \citealt{Connolly},
\citealt{Steidel}, \citealt{Treyer}, \citealt{Sullivan},
\citealt{GWilson}), H$\alpha$
fluorescent emission (\citealt{Gallego}, \citealt{Tresse},
\citealt{Glazebrook}, \citealt{Sullivan}), radio emission
(\citealt{Serjeant}, \citealt{Haarsma}) and far-infrared emission
\citep{Flores}.  Most of the estimates of the SFR are convergent to $\approx
0.2$ dex among different authors, if the same observational techniques
are used. The large uncertainty, however, resides in which techniques
are to be used; Current estimates show an uncertainty of a factor of
$\approx 6$ (0.8 dex). In particular, the SFR estimated from UV depends
largely on extinction corrections. \citet{Madau-etal} took
$E_{B-V}=0.1$, \citet{Steidel} indicated $0.15$ and \citet{Sullivan}
derived $0.13$. This corresponds to an uncertainty of 0.36 dex in the
SFR. The prime purpose of this paper is to consider whether we can
obtain any meaningful constraints on the SFR from the current
observation of the supernova relic neutrinos.

Another focus of this paper is to derive a lower limit on the relic
neutrino event rate expected in the SK detector under reasonable
assumptions on the input to the calculation.  The calculation of the
event rate suffers from uncertainties besides the SFR.  Among the most
important uncertainties are those in the spectrum of neutrinos.  In this
paper we try to reduce the uncertainty in the neutrino spectrum using
observation of neutrinos from SN1987A as a constraint. We may assume
that it is typical of type II supernovae, since we expect that the
physics of core collapse is similar even if the optical appearance may
have a wide variety. The other uncertainty concerns neutrino
oscillation, in which $\nu_e$ and $\nu_\mu$ partly interchange during
propagation through vacuum and Earth. Recent neutrino oscillation
experiment showed unambiguously the presence of neutrino oscillation,
and determined the oscillation parameters (\citealt{SK1}, \citealt{SNO},
\citealt{SK2}). We can now calculate accurately the effect of
oscillation both in vacuo and in Earth. This is no longer a source of
uncertainties.

Throughout this paper we adopt the natural units, $c=\hbar=1$, and the
Boltzmann constant $k_B=1$.

\section{The local supernova rate and the local star formation rate}

A number of extragalactic supernova surveys (\citealt{Bergh},
\citealt{Cappellaro}, \citealt{Tammann}) yield the local supernova rate
in units of SNu, {\it i.e.}, the number of supernovae per
$10^{10}L_B(\odot)$ per 100 year for each morphological type of
galaxies. We translate it into the rate per unit cube of spatial volume,
by averaging over morphological fractions of nearby galaxies E/S0 :
Sa-Sb : Sbc-Sd = 0.32 : 0.28 : (\citealt{Fukugita}; hereafter FHP), and
by multiplying the $B$ band local luminosity density of the universe
${\cal L}_B=2.4\pm0.4\times 10^{8}hL_\odot$ Mpc$^{-3}$
(\citealt{Blanton}; \citealt{Yasuda}), where $h$ is the Hubble constant
in units of 100 km s$^{-1}$Mpc$^{-1}$. Counting both type Ib and Ic in
addition to type II as core-collapse supernovae, we obtain the supernova
rate $R_{\rm SN}$ as $1.98\times 10^{-4}$, $2.11 \times 10^{-4}$ and
$4.44 \times 10^{-4}$ $h^3$yr$^{-1}$Mpc$^{-3}$ for the three surveys. We
take the geometric mean of the three values and refer to the largest and
smallest of the three as the allowed range: $2.65{+1.79 \atop
-0.67}\times 10^{-4}$ $h^3$yr$^{-1}$Mpc$^{-3}$. This is compared with a
similar estimate $4.7\times 10^{-4}$$h^3$yr$^{-1}$Mpc$^{-3}$
by~\citet{Madau-etal}.

For a given star formation rate $\psi(t)$ (in units of  M$_{\odot}$
yr$^{-1}$Mpc$^{-3}$) $R_{\rm SN}$ is calculated as
\begin{equation}
      R_{\rm SN}=\psi(t) \frac{\int^{m_u}_{m_c} dm \phi(m)/m}
      {\int^{m_u}_0 dm \phi(m)},
      \label{eq:sn-rate}
\end{equation}
where $\phi(m)$ is the initial mass function (IMF) of stars, for which
we take the Salpeter form $\phi(m)\sim m^{-1.35}$ for $m>1M_\odot$ and
continue to the IMF of \citet{Gould} for a low mass (see FHP) so that
the integral in the denominator is extended to zero mass. We set
$m_u=100M_\odot$. Since the Salpeter IMF has been adopted in virtually
all literature that derived SFRs, our calculation of SFR can be directly
compared to those, only with a downward correction of a factor of 1.65
that arises from the difference between the genuine Salpeter IMF with a
lower cut off of 0.1$M_\odot$ and our prescription.  We take the
critical mass for type II supernovae $m_c$ to be between 8 and 10
$M_\odot$ according to \citet{Nomoto}.

The local star formation rate derived from (\ref{eq:sn-rate}) is
$\log~\psi(t_0)( M_{\odot}$yr$^{-1}$Mpc$^{-3}$)$ = -2.09 {+0.22 \atop
-0.13}$, where we take $h=0.72$. This may be compared with the SFR from
an H$\alpha$ survey $-2.11\pm0.04$ (\citealt{Tresse},
\citealt{Glazebrook}; we take their independent data points).

\section{Star formation rate and the relic neutrino event rate in
the detector}

\begin{figure}
      \centering \includegraphics[width=8cm]{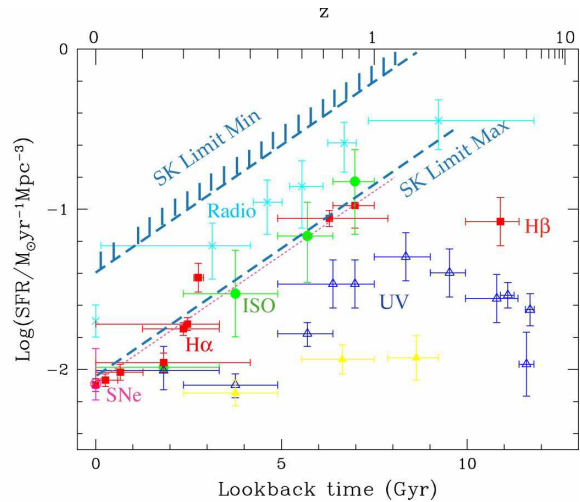}
      \caption{Star formation rate (SFR) inferred from H$\alpha (\beta)$
      (filled squares), UV emissivity (open and filled triangles),
      far-infrared (filled circles) and radio (crosses) observations as a
      function of lookback time. The SFR estimated from the local supernova
      rate is shown by open circle at 0 Gyr.  The two dashed lines denote
      the 90\% SK limits for the case of the minimum event rate (the
      conservative limit; hatching attached) and for the case of maximum
      event rate within uncertainties of the model supernova neutrino
      flux. The dotted line is fit~(\ref{eq:Halpha-fit}).  All data use
      the modified Salpeter IMF described in the text.}
      \label{fig:sfr}
\end{figure}

\smallskip
\noindent
{\it Star formation rates}
\smallskip

In Figure~\ref{fig:sfr} we present estimates for the SFR as a function
of the lookback time ($= t_0-t$ with $t_0$ the present time, in units of
Gyr). The data are taken from~\citet{Glazebrook}, \citet{Sullivan},
\citet{Tresse}, \citet{Steidel} and \citet{Haarsma}, which cover most of
the SFR work to date
We include in the Figure the UV estimates of \citet{GWilson} (filled 
triangles with thin drawings). These data, however, largely disagree 
with other estimates and the reasons are unknown: so we do not
refer to them further in this paper. 
We take the cosmology of $\Omega_0=0.3$ and $\lambda=0.7$ with $h=0.72$
to draw this figure. The solid points are the SFR from H$\alpha$ (and
H$\beta$) with the extinction estimated using the Balmer decrement, and
the open points refer to the values from UV emissivity assuming zero
extinction corrections. Radio observations are denoted by crosses. The
grey points are obtained from far-infrared (ISO) observations. The SFR
estimate from the supernova rate is shown at the zero lookback time. We
also show the constraint from the supernova relic neutrino observations
obtained from the present calculation given below.

The figure shows that the SFR obtained from a single indicator exhibits
nearly an exponential dependence as a function of the lookback time, at
least for $z\leq 1$, as expected in the closed box model and also in CDM
model calculations (e.g. \citealt{Nagamine}). For example the SFR for
$t-t_0 < 7$~Gyr obtained from H$\alpha$ is fitted well with
\begin{equation}
      \label{eq:Halpha-fit}
      \log \psi(t)_{{\rm H}\alpha}
       = (-1.96\pm0.04) + \log~h + 0.216 h (t_0- t).
\end{equation}
The SFR derived from far-infrared observations is consistent with this
curve. The radio data give a consistent slope, but the normalisation is 
dex
times higher.  The data from UV emissivity also indicate a line parallel
to (\ref{eq:Halpha-fit}), but located lower by 0.50 dex With the
standard extinction law (\citealt{Seaton}, \citealt{Cardelli})
$E_{B-V}=0.20\pm 0.05$ brings the UV data consistent with the H$\alpha$
data. In the following argument we assume the exponential law for the
SFR
\begin{equation}
      \psi(t)=\psi(t_0)\exp[(t_0-t)/\tau]
      \label{eq:sfr}
\end{equation}
with $\tau= 2.8$ Gyr for $z<1$. This law may not hold for $z>1$, but
star formation at such high redshift is insignificant in a estimate of
the supernova relic neutrino flux for $E_\nu>12$~MeV, which is of our
current concern, and we do not need to specify any accurate functional
form, as we will see below. We take $\psi(t_0)$ as a
parameter. Alternatively, we may use $E_{B-V}$ as a parameter taking
$\log \psi_{\rm UV}(t_0)=-2.45$ from UV emissivity with zero extinction
as a fiducial value.

\medskip
\noindent
{\it Neutrino spectrum from type II supernovae}

We must deal with the uncertainty of the neutrino flux emergent from
type II supernovae. The dominant neutrino emission arises from pair
creation in the optically thick object. Hence the luminosity of each
species of neutrinos is approximately equal, as demonstrated by many
neutrino transport calculations given in Table~\ref{tab:ave}. The total
neutrino luminosity is close to $3\times 10^{53}$ erg, since the mass of
all observed neutron stars takes a universal value of 1.4$M_\odot$. We
allot a 20\% error to this estimate. The mean energy of neutrinos
depends on details of calculations, {\it e.g.} ranging from 12 to 20 MeV
for $\bar\nu_e$.  The mean neutrino energies satisfy $\langle
E_{\nu_e}\rangle <\langle E_{\bar\nu_e}\rangle < \langle
E_{\nu_{\mu\tau}}\rangle$, where $\nu_{\mu\tau}$ includes $\nu_\mu$,
$\nu_\tau$ and their antiparticles. Neutrino transport calculations give
$E_{\nu\mu}/E_{\bar\nu_e}= 1.6$; see Table~\ref{tab:ave}. We take this
ratio to be 4/3, but the result of our calculation is not sensitive to
this ratio once we introduce the observational constraint.

The model neutrino spectrum from supernovae is somewhat deviated from
the zero-chemical potential Fermi distribution. It is usually
parametrised by introducing an effective chemical potential $\eta$,
i.e., $ f=[\exp{(E_\nu/T_\nu-\eta)}+1]^{-1}$ with $\eta=1-3$
(\citealt{Janka}).

For water $\check{\rm C}$erenkov neutrino detectors the only reaction we
must consider is $\bar\nu_e p\rightarrow e^+n$. The cross section of
$\nu_e ^{16}$O is $>20$ times smaller. There is, however, an important
contribution from $\bar\nu_\mu\rightarrow\bar\nu_e$ due to neutrino
oscillation.  Recent solar neutrino experiments at Sudbury and SK show
that the mixing between $\nu_e$ and $\nu_\mu$ is nearly maximal.  With
the matter effect the neutrinos emergent from a supernova are
$\nu_2=-\sin\theta\nu_e+\cos\theta\nu_\mu$ and $\nu_1=
\cos\theta\nu_e+\sin\theta\nu_\mu$, which are the mass eigenstates. The
$\bar\nu_e$ detected in detectors are therefore $\cos^2\theta$ times the
$\bar\nu_e$ flux and $\sin^2\theta$ times the $\bar\nu_\mu$ flux where
$\sin^22\theta\simeq0.96$ (\citealt{SNO}, \citealt{SK2}). This in
principle increases the neutrino detection rate due to higher energies
of the $\bar\nu_\mu$ flux.

\begin{table}
      \centering
      \caption{Mean energies (in units of MeV) of supernova neutrinos.}
      \begin{tabular}{|l|c|c|c|}  \hline
         &  $\langle E_{\bar{\nu}_e} \rangle$
          & $\langle E_{\bar{\nu}_{\mu}} \rangle$
          &  $\langle E_{\bar{\nu}_{\mu}} \rangle
               /\langle E_{\bar{\nu}_{e}} \rangle$\\ \hline
          \citet{Burrows}  &   $12$    &  $22$   &  $1.8$ \\
          \citet{Bruenn}   &   $21$    &  $24$   &  $1.1$ \\
          \citet{Wilson}   &   $15$    &  $20$   &  $1.3$ \\
          \citet{Yamada}   &   $16$    &  $24$   &  $1.5$ \\ \hline
      \end{tabular}
      \label{tab:ave}
\end{table}

The matter effect of Earth somewhat modifies the mixing
ratio for the neutrino flux that passes through Earth. This effect
is calculated assuming Earth as a sphere of a constant
matter density, $\rho_e({\rm Earth}) \simeq  3.2$g/cm$^3$.

\medskip
\noindent
{\it Constraints from SN1987A}

At the epoch of SN1987A Kamiokande~\citep{Hirata} and
Irvine-Michigan-Brookhaven Collaboration (IMB)~\citep{Bionta} detected
neutrinos from core collapse. The gross characteristics of these
neutrino events agree with what are expected. Here we use the detection
of neutrino events at IMB, which has a larger fiducial volume and is
more sensitive to rare, higher-energy neutrino events, to constrain the
higher energy spectrum of supernova neutrinos. We estimate the event
number $N_{\rm IMB}$ at IMB (5 kton water) as

\begin{equation}
      N_{\rm IMB} = 3.3\times 10^{32}
      \int_{20 {\rm MeV}}^{60 {\rm MeV}} dE_\nu
      \sigma_p(E_\nu)F_{\bar{\nu}_e}(E_\nu)x(E_\nu),
\end{equation}
where $\sigma_p(E_\nu)$ is the cross section for $\bar\nu_ep\rightarrow
e^+n$ with the neutrino energy $E_\nu$, $x(E_\nu)$ is the trigger
efficiency, and $F_{\bar{\nu}_e}$ is the $\bar{\nu}_e$ flux at the IMB
detector,
\begin{equation}
      \label{eq:neutrino-flux}
      F_{\bar{\nu}_e}(E_\nu) =
      P_{\rm IMB}(E_\nu){\cal F}_{\bar{\nu}_e}
      +(1-P_{\rm IMB}(E_\nu)){\cal F}_{\bar{\nu}_{\mu}},
\end{equation}
where  $P_{\rm IMB}(E_\nu)$ is the
conversion probability for $\bar{\nu}_1 \rightarrow \bar{\nu}_{e}$
due to neutrino oscillation including matter effects of Earth and
${\cal F}_i$ is the neutrino flux for species $i$ without
oscillation, for which we obtain
\begin{eqnarray}
      {\cal F}_i  & = & 3.99\times 10^{10} {\rm cm}^{-2}{\rm MeV}^{-1}
      \left(\frac{E_\nu}{\rm MeV}\right) \nonumber \\
      & & \times \left(\frac{T_{\nu_i}}{\rm MeV}\right)
      \left(\frac{E_{\rm tot,\nu_i}}{10^{53}{\rm erg}}\right)
      C(\eta)f(\eta, E_\nu),
\end{eqnarray}
where $E_{\rm tot,\nu}$ is the total neutrino energy, $T_{\nu_i}$ is
temperature of neutrino for $i$th species and $C(\eta) = \int dE_\nu
E_\nu^3 f(\eta, E_\nu)/ \int dE_\nu E_\nu^3 f(0, E_\nu)$.  The
calculation of $P_{\rm IMB}$ is standard and is carried out in a way
similar to that in \citet{Smirnov}, taking neutrino trajectory inside
Earth for SN1987A. Applying Poisson statistics to $8$ events observed at
IMB, we obtain $90$\% confidence limits on $\langle E_{\bar{\nu}_e}
\rangle$ and $T_{\bar{\nu}_e}$ as shown in Table~\ref{tab:IMB}. This
constraint removes much of the uncertainty of the model neutrino flux:
hence the result of our calculation in what follows depends only weakly
on the assumptions on parameters we assumed for the model neutrino flux
emergent from supernovae.

\begin{table}
      \centering
      \caption{90\% confidence allowed range of  the neutrino energy and
      temperature from IMB neutrino events for SN1987A.}
      \begin{tabular}{|c|c|c|}  \hline
        $\eta$ &  $\langle E_{\bar{\nu}_e} \rangle$ (MeV)
          & $T_{\bar{\nu}_e}$ (MeV)\\ \hline
          0   &   $10.5 - 14.9$    &  $3.33 - 4.72$ \\
          1   &   $11.0 - 15.3$    &  $3.31 - 4.60$ \\
          2   &   $11.4 - 15.8$    &  $3.16 - 4.39$ \\
          3   &   $12.0 - 16.4$    &  $3.01 - 4.11$ \\ \hline
      \end{tabular}
      \label{tab:IMB}
\end{table}

\medskip
\noindent
{\it Supernova relic neutrino flux}
\medskip

The neutrino flux ${\cal J}_{\nu_i}$ is calculated as
\begin{equation}
      \label{eq:relic-flux}
      {\cal J}_{\nu_i}(E_\nu) = \int_{0}^{z_f} dz
      \frac{-dt}{dz} (1+z)
      {\cal L}_{\nu_i}((1+z)E_\nu)R_{\rm SN}(z),
\end{equation}
where the neutrino luminosity ${\cal L}_{\nu}$ is given by
\begin{equation}
      {\cal L}_{\nu}  =E_{\rm tot, \nu}\frac{120}{7\pi^4}
      \frac{E_\nu^2}{T_{\nu}^4}C(\eta)f(\eta,E_\nu).
\end{equation}
$z_f$ is the formation epoch of galaxies, but the integral is dominated
by a low redshift region. The supernova rate is related with the SFR as,

\begin{equation}
      R_{\rm SN}(z)  =  1.22 \times 10^{-2}\psi(z)/M_\odot
\end{equation}
for $m_c=8M_{\odot}$. We remark that the flux (\ref{eq:relic-flux}) is
independent of the cosmology, since its dependence in $dt/dz$ is
compensated by the volume factor of $\psi(z)$.  We use for $\psi$ the
exponential law for $z<1$ and assume a constant for $z>1$. We take
$\psi(t_0)$ as a free parameter. This parameter is also translated to
$E_{B-V}$, taking the SFR from UV emissivity with zero extinction as a
fiducial. The extinction is written $\Delta \log\psi=2.42 E_{B-V}$ for
$z<1$ using the standard extinction law (\citealt{Seaton},
\citealt{Cardelli}).

\begin{figure}
      \centering
      \includegraphics[width=8.5cm]{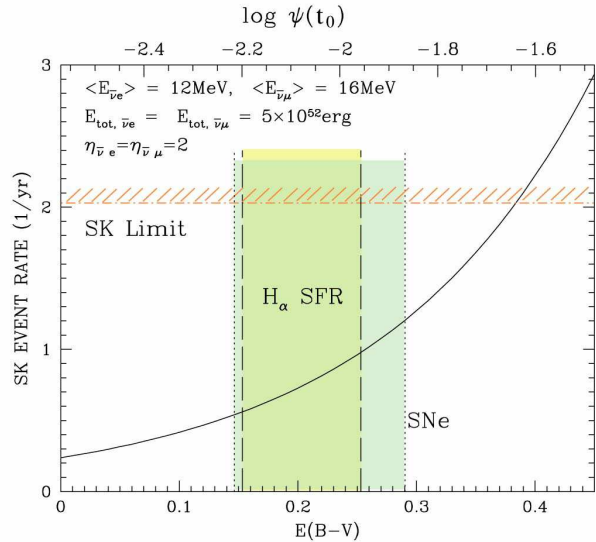}
      \caption{Neutrino event rate for SK as a function of $E_{B-V}$. The
      hatch is the limit from SK. The range between the two dashed lines
      is consistent with SFR from H$\alpha$ observations, and that
      between the two dotted lines is consistent with the local supernova
      rate.} \label{fig:event-ext}
\end{figure}

The calculation of supernova relic neutrino flux $J_{\bar{\nu}_e}$
(after neutrino oscillation effect) is carried out in a similar way,
with the modification that the Earth effect is integrated over all
directions. The event rate with a fiducial volume of the 22.5~kton SK
detector is calculated as
\begin{eqnarray}
      R_{\rm SK}  & = &
      4.2\times 10^{-3}{\rm yr}^{-1} \nonumber \\
      & & \times \int_{E_{\nu_{\rm min}}}^{E_{\nu_{\rm max}}}
      \left(\frac{E_\nu}{\rm MeV}\right)^2
      \left(\frac{J_{\bar{\nu}_e}(E_\nu)}
        {{\rm MeV}^{-1}{\rm cm}^{-2}\sec^{-1}}
      \right),
\end{eqnarray}
where $E_{\nu_{\rm min(max)}}$ is the minimum (maximum) energy of the
neutrino detection, which we take $E_{\nu_{\rm min}}=18$~MeV and
$E_{\nu_{\rm max}}=50$~MeV. Fig.~\ref{fig:event-ext} shows an example
result for the neutrino event rate for SK as a function of $E_{B-V}$ for
a typical parameter set indicated in the figure. It is seen that the
event rate exceeds the current SK limit, $2.0$~yr$^{-1}$ at 90\%
confidence for the $18-50$~MeV window~\citep{Totsuka}, if $E_{B-V}\gsim
0.4$.  From this figure we can read the range of $E_{B-V}$ which is
consistent with the SFR from H$\alpha$ at the zero redshift. The SFRs
 from H$\alpha$ and the local supernova rate are consistent.

\begin{figure}
      \centering
      \includegraphics[width=8.5cm]{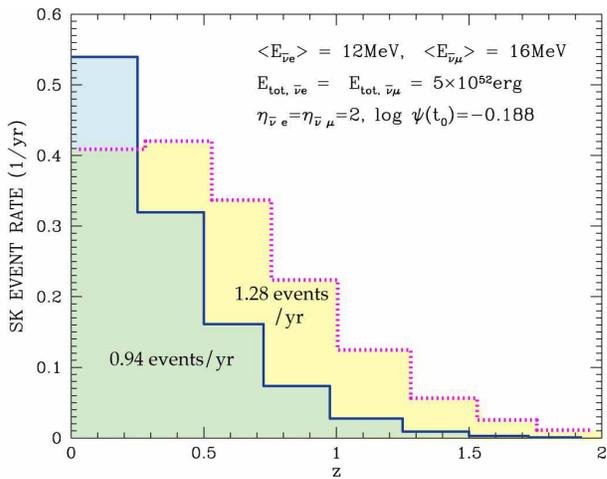}
      \caption{Neutrino events partitioned according to the redshift of
      supernovae. The solid (dotted) histogram corresponds to the
      energy window of 18$-$50 MeV (12$-$18MeV).
      We take the parameter set the same as that
       in Fig.~\ref{fig:event-ext}.}
      \label{fig:event-redshift}
\end{figure}

\begin{figure}
      \centering
      \includegraphics[width=8.5cm]{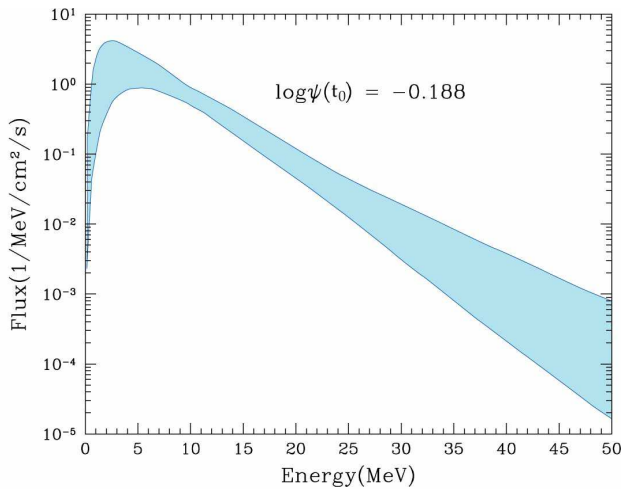}
      \caption{Neutrino spectrum expected at the detector after neutrino
      oscillation. Uncertainties in the spectral shape are represented
      by shading.}
      \label{fig:spectrum}
\end{figure}

Fig.~\ref{fig:event-redshift} shows the relative importance of
supernovae at different redshifts for neutrino events. The solid
histogram corresponds to neutrino events with energy between 18~MeV and
50~MeV.  It shows that half the events arise from low-$z$ supernovae
($z<0.25$). If we decrease the energy of the detector window to
$12<E<18$ MeV, non-zero redshift supernovae become more important. This
means that we can learn the evolution of the SFR from gross spectroscopy
of supernova relic neutrinos. This histogram also shows that the
contribution from supernovae at $z>1$ is insignificant in so far as our
consideration is restricted to $E>12$ MeV. The neutrino spectrum is
presented in Figure~\ref{fig:spectrum}.

\medskip
\noindent
{\it Constraint on the star formation rate}
\medskip

We derive a constraint on the SFR by requiring that the supernova relic
neutrino event should not exceed the SK limit allowing for uncertainty
of the model neutrino flux shown in Fig.~\ref{fig:event-ext}; the result
is shown in Figure~\ref{fig:sfr} above. The most conservative limit is
obtained by taking $m_c=10M_{\odot}$ and minimising the number of events
at IMB for SN1987A. At a 90\% confidence the limit means
\begin{equation}
      \label{eq:const-ext}
      \log\psi(t_0)  \le  -1.40, ~~~~{\rm or}~~~~E_{B-V}  \le   0.48.
\end{equation}
The constraint derived here is about 5 times higher than the SFR
obtained from H$\alpha$ and far-infrared, and twice higher than the SFR
 from radio observations.

If we would take the opposite case, {\it i.e.} the maximum allowed flux
(90\% confidence for the IMB events) and $m_c=8M_{\odot}$, the limit
becomes stronger by a factor of 4; it nearly coincides with the SFR from
H$\alpha$, and already overshoots the SFR from radio.

We emphasize that large uncertainties in the model supernova neutrino
flux calculations are significantly reduced by empirical constraints
derived from SN1987A.  For instance an increase of $\eta$ is compensated
by an increase of effective neutrino energy or else by an increase of
neutrino luminosity, so that the neutrino flux in the energy range that
concerns us changes little. Neutrino oscillation generically enhances
the high energy tail of the neutrino spectrum. Under the empirical
constraint, however, this is absorbed into the change of other
parameters. As a result the prediction of supernova relic neutrinos at
SK is modified little.

\medskip
\noindent
{\it The lower limit of neutrino reaction rates at SK}

Assuming $\log\psi=-2.15$, which is the lower value of SFR from
H$\alpha$ allowed within the error range (or SFR from UV with $E_{B-V}
\ge 0.19$) and taking our minimum neutrino flux estimate and
$m_c=10M_\odot$, we obtain

\begin{equation}
     R_{\rm SK} \ge 0.40~{\rm yr}^{-1}.
\end{equation}
A similar limit is derived if we take the lowest value of the
supernova rate:
\begin{equation}
      R_{\rm SK} \ge  0.44~{\rm yr}^{-1}.
\end{equation}
The SK should see the supernova relic neutrino events if
they increase the sensitivity by a factor of 5.

\section{Conclusions}

The limit on the SFR derived from the supernova relic neutrino
observation at SK is already marginally significant even if we include
the uncertainty of the model supernova neutrino flux. With the current
data we can conclude that $\psi(t_0)<0.040
M_{\odot}$yr$^{-1}$Mpc$^{-3}$, which is 5 times the estimate from
H$\alpha$ (0.0078$\pm0.0008M_{\odot}$yr$^{-1}$Mpc$^{-3}$) and twice that
 from radio observations. The SFR from local supernova surveys
(0.0081${+0.0054\atop-0.0021}M_{\odot}$yr$^{-1}$Mpc$^{-3}$) is
consistent with the estimate from H$\alpha$. For the SFR from UV
emissivity, our result means $\langle E_{B-V}\rangle<0.48$ with the
standard extinction law. The increase of SK statistics by a factor of 5
should positively detect the supernova relic neutrino events.

\smallskip
\noindent
{\bf ACKNOWLEDGMENTS}
\smallskip

We would like to thank Yoji Totsuka and Ken Nomoto for valuable
discussions.

\end{document}